\documentclass[runningheads]{llncs}
\usepackage{wrapfig,lipsum,booktabs}
\usepackage{amsfonts,amsmath,amssymb,amstext}
\usepackage[ruled,vlined,linesnumbered]{algorithm2e}
\usepackage[normalem]{ulem}
\usepackage{graphicx,color}
\usepackage{subfig}

\newcommand{\comment}[1]{}

\newcommand{\fullOnly}[1]{}

\newcommand\f\varphi
\newcommand\Id\varepsilon

\usepackage{setspace}

\SetKwInOut{Input}{Input}
\SetKwInOut{Output}{Output}
\SetKwInOut{GV}{Global}
\SetKwProg{Fn}{Function}{}{}
\SetKwProg{Proc}{Procedure}{}{}
\DontPrintSemicolon

\newcommand\donotshow[1]{}

\newcommand{\shapeit}{\textsc{ShapeIt}}
\newcommand{\shshape}{\textsf{shape}}
\newcommand{\aline}{\textsf{line}}
\newcommand{\shunion}{\textbf{+}}
\newcommand{\shstar}{\textbf{*}}
\newcommand{\shconcat}{\textbf{.}}
\newcommand{\shconstraint}{\textsf{cst}}
\newcommand{\shcolumn}{\textbf{:}}
\newcommand{\SE}{\textsf{SE}}
\newcommand{\shand}{\textsf{and}}

\begin{document}

	\title{Mining Shape Expressions with ShapeIt\thanks{This  project  has  received  funding  from  the  European Union’s  Horizon  2020  research  and  innovation  programme  under  grant  agreement No 956123 and it is partially funded by the TU Wien-funded Doctoral College for SecInt: Secure and Intelligent Human-Centric Digital Technologies.}}
	\titlerunning{Mining Shape Expressions with ShapeIt}
	\author{
		Ezio Bartocci\inst{1} \and
		Jyotirmoy Deshmukh\inst{2} \and
		Cristinel Mateis\inst{3} \and \\
		Eleonora Nesterini\inst{1,3} \and
		Dejan Ni\v{c}kovi\'{c}\inst{3} \and
		Xin Qin\inst{2}
	}
	%
	%
	\institute{
		TU Wien, Austria
		\and
		University of Southern California, USA
		\and
		AIT Austrian Institute of Technology, Austria
	}

	\maketitle              
	\begin{abstract}


We present \textsc{ShapeIt}, a tool for mining specifications of cyber-physical systems (CPS) from their real-valued behaviors. 
The learned specifications are in the form of {\em linear shape expressions}, 
a declarative formal specification language suitable to express behavioral properties 
over real-valued signals. A linear shape expression is a regular expression composed of parameterized lines as atomic symbols with symbolic 
constraints on the line parameters. We present here the architecture of our tool
along with the different steps of the specification mining algorithm. We also  describe the usage of the tool demonstrating its applicability on 
several case studies from different application domains.

	\end{abstract}

	\section{Introduction}

Specification mining~\cite{NenziSBB18,BBS14,Wang2020} is the process of inferring likely system properties from observing its execution and the behavior of its environment. This is an emerging research field that supports the engineering of cyber-physical systems (CPS) where computational units are tightly embedded with physical entities such as sensors and actuators controlling a physical process. CPS often operate (autonomously) in sophisticated and unpredictable environments. 

In this context, mined properties can be used
to complete existing incomplete or outdated specifications, to understand essential properties of black-box components (e.g., machine learning components) and to automate difficult tasks such as fault-localization~\cite{BartocciFMN18,JinDDS15}, failure explanation~\cite{BartocciMMMN19} and
falsification analysis~\cite{BartocciDDFMNS18}. The symbolic and declarative nature of formal specification languages provide an high-level and abstract framework that facilitates generalisation. Furthermore, mined specifications are re-usable, data-efficient, compositional and closer to human understanding.

In this paper, we present \textsc{ShapeIt}, a tool for automatic mining formal specifications from positive examples of time-series data encoding system behaviors or a discrete-time trace of the value of a particular system variable.  \textsc{ShapeIt} uses {\em Linear Shape Expressions} (LSEs)~\cite{NickovicQFMD19}, a recent introduced declarative formalism  suitable  to  express  expected behaviors over noisy real-valued signals. A linear shape expression is a regular expression composed of parameterized lines as atomic symbols with symbolic constraints on the line parameters.

Given a set of time-series and a maximum error threshold, 
\textsc{ShapeIt} implements the specification mining procedure~\cite{BartocciDGMNQ20} 
consisting of three steps: (1) \textbf{segmentation} of 
time-series into an optimal piecewise-linear approximation, 
(2) \textbf{abstraction} and \textbf{clustering} of linear segments 
into a finite set of symbols, where each symbol represent a set of 
similar lines, and (3) \textbf{learning} of linear shape expressions 
from the sequences of symbols generated in the previous step.

In the rest of the paper, we present the specification language and the 
architecture of the tool. We also show
the usage of our tool, demonstrating the applicability to 
several different examples of time-series taken from the literature.
The code of our tool is publicly available at: {\small\url{https://www.doi.org/10.5281/zenodo.5569447}}.
\vspace{-2ex}

	\section{Shape Expressions}
\label{sec:seshort}

{\em Linear shape expressions} (LSE)~\cite{NickovicQFMD19} are regular expressions defined 
over parameterized {\em linear atomic shapes}, where a linear atomic shape is uniquely 
determined by three parameters: slope $a$, (relative) offset $b$ and duration $d$. 
LSEs can have additional constraints over these parameters.
We use the following syntax to define the fragment of LSEs 
supported by $\shapeit$.

$$
\begin{array}{lcl}
\shshape & := & \aline(a,b,d)~|~\shshape_{1} \; \shunion \; \shshape_{2}~|~\shshape_{1} \; \shconcat \; \shshape_{2}~|~(\shshape) \shstar \\
\shconstraint & := & x \; \textsf{in} \; \textbf{[c{1}, c{2}]}~|~\shconstraint_{1} \; \shand \; \shconstraint_{2} \\
\SE & := & \shshape \; \shcolumn \; \shconstraint \\
\end{array}
$$
\noindent where $\textbf{c1}$ and $\textbf{c2}$ are rational constants such that 
$\textbf{c1} \leq \textbf{c2}$.

A LSE $\SE$ consists of two main components, a regular expression $\shshape$ that 
captures the qualitative aspect of the specification,
and a constraint $\shconstraint$ imposed on the LSE parameters.
Shape expressions are evaluated against finite signals -- sequences of 
(time, value) pairs. The semantics of LSE is defined in terms 
of a {\em noisy match} relation. We say that a signal is a 
$\nu$-noisy match of a linear atomic shape, if there exists an ideal line segment 
with some slope $a$, relative offset $b$ and duration $d$ such that (1) 
$a$, $b$ and $d$ satisfy the constraint $\shconstraint$, and (2) the 
mean square error (MSE) between the signal segment and the ideal line segment 
is smaller than or equal to $\nu$. This definition is inductively lifted to arbitrary 
LSEs. In essence, a signal is a $\nu$-noisy match of an arbitrary LSE if 
there exists a sequence of linear atomic shapes with instantiated parameters such 
that: (1) the sequence is consistent with the qualitative (regular expression) 
part of the LSE, (2) the instantiated parameters satisfy the LSE constraint, and 
(3) the signal can be split into the sequence with the same number of segments, 
such that each signal segment is a $\nu$-noisy match of its corresponding 
atomic shape. The formal syntax and semantics of 
shape expressions are presented in~\cite{BartocciDGMNQ20}.

	\section{ShapeIt Architecture, Methods and Implementation}
\label{se:arch}

\begin{figure}[htp]
	    \centering
	     {\includegraphics[width=0.8\textwidth]{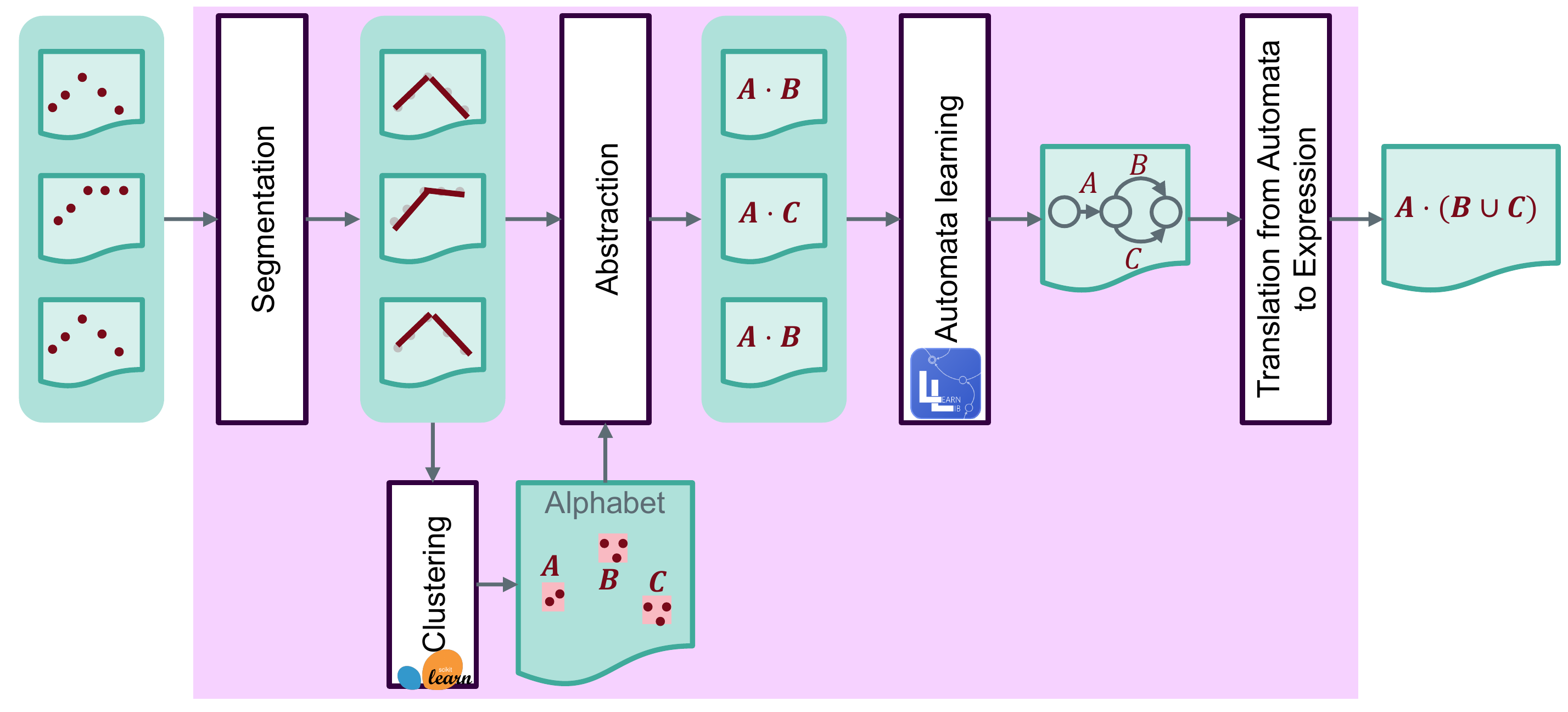}}
	    \caption{Overview of \textsc{ShapeIt} workflow.\label{fig:overview}}
	    \vspace{-5ex}
	    \end{figure}

The architecture of \textsc{ShapeIt} is depicted in Figure~\ref{fig:overview}. 
The tool consists of five components: (1) segmentation, (2) abstraction, (3) clustering, 
(4) automata learning and (5) translation from automata to regular expressions. 
\textsc{ShapeIt} is implemented in Python 3 with the use of external Python and 
Java libraries.

\paragraph{Segmentation} module implements the piecewise-linear approximation 
algorithm with quadratic complexity from~\cite{BartocciDGMNQ20} 
that given a time series and a mean square error (MSE) threshold 
computes the minimal sequence of segments such that for each segment of 
data, its linear regression MSE is below the threshold. The input of this 
module is a set of time-series and the output is a set of line segment sequences, 
where each line segment is characterized by slope, relative offset and duration 
parameters.

\emph{Abstraction and clustering} module takes as input the 
set of line segments (computed by the segmentation module) and uses 
the k-Means clustering implementation from the \textsf{scikit-learn} 
library\footnote{\url{https://scikit-learn.org/stable/}} 
to group lines with similar parameters. The user specifies 
a threshold on the derivative of the Within-Cluster-Sum-of-Squares (WCSS) 
error measure to determine the optimal number of clusters. The tool 
defines a finite alphabet in which each letter is
associated to a different cluster. Each letter is also assigned the 
minimal bounding cube that contains all the points in 
its corresponding cluster. Each line segment is mapped to a letter in 
the alphabet, resulting in a set of finite words.

\emph{Automata learning} module applies the Regular Positive
and Negative Inference (RPNI) algorithm for passive learning from 
positive examples, implemented in the Java \textsf{learnlib} 
library\footnote{\url{https://learnlib.de/}}, 
to infer a deterministic finite automaton (DFA) from a set of finite words. 
The integration of the Java library in our Python implementation is done 
using the JPype library.\footnote{\url{https://jpype.readthedocs.io/en/latest/}}

\emph{DFA to shape expressions} module implements the algorithm 
for translating DFAs to regular expressions using the state elimination method. 
The NetworkX library\footnote{\url{https://networkx.org/}} is used to represent and manipulate DFAs during the 
translation.

    \section{Evaluation}
In the following, we evaluate the applicability of \textsc{ShapeIt}\footnote{commit in the repository used: d92341998d66615cf6a9c4f3bcc419df4cd988b6} to find temporal patterns over different time-series datasets stored in the UCR Time Series Classification Archive~\cite{UCRArchive}.  Our experiments run on a Notebook Dell Latitude 5320, Intel Quad-Core i7-1185G7 (3,00 GHz/Turbo 4,80 GHz), RAM 32 GB. \textsc{ShapeIt} software components run on Python version 3.8.8 and on Java version 16.0.2.  For all the experiments we set to 10 the threshold on the derivative of the WCSS error discussed in Section~\ref{se:arch}.

\vspace{-2ex}
\begin{figure}[htp]
	    \centering
	    \includegraphics[width=\textwidth]{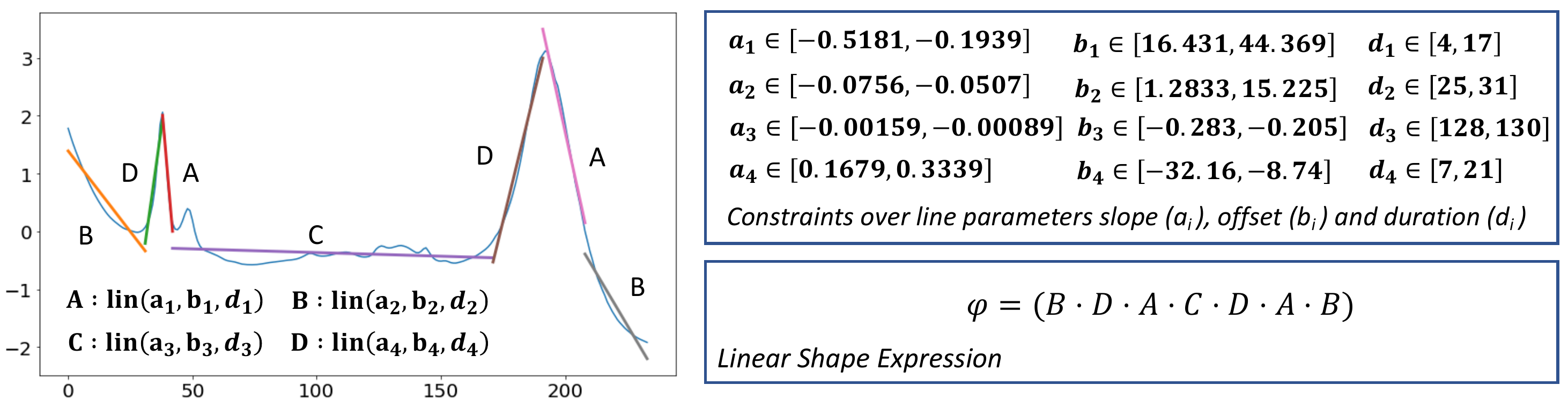}
	    \caption{(Left) An example of piece-wise linear approximation of a trace in \textit{Wine dataset} with $\varepsilon_{\max} = 0.05$ (Right) Generated Linear Shape Expression.}
	    \label{fig:wine}
	    \vspace{-7ex}
	\end{figure}
	
\begin{wraptable}{r}{8.5cm}
\begin{tabular}{|cc|cccccc|cc|}
\toprule \scriptsize{ \# traces} &
  \scriptsize{$|w|$}  &&  \scriptsize{ $t_s(s)$}    &&  \scriptsize{$t_c(s)$} && \scriptsize{$t_l(s)$} && \scriptsize{$t_{\text{total}} (s)$}      \\

\midrule
  \scriptsize{$1$} & \scriptsize{$10$} && \scriptsize{$2.205 \cdot 10^{-4}$} &&  \scriptsize{$1.100 \cdot 10^{-6}$}   &&  \scriptsize{$3.499 \cdot 10^{-4}$} && \scriptsize{$5.724 \cdot 10^{-4}$}\\
 
 \scriptsize{$1$} & \scriptsize{$100$} && \scriptsize{$7.173 \cdot 10^{-2} $}  && \scriptsize{$ 4.968 \cdot 10^{-3}$}  && \scriptsize{$ 3.727 \cdot 10^{-4} $} && \scriptsize{$ 7.707 \cdot 10^{-2} $} \\
 
 \scriptsize{$1$} & \scriptsize{$234$} && \scriptsize{$4.227 \cdot 10^{-1}$} && \scriptsize{$5.195 \cdot 10^{-3}$}  && \scriptsize{$ 4.319 \cdot 10^{-4} $} && \scriptsize{$4.283 \cdot 10^{-1}$}\\
 
 \scriptsize{$10$} & \scriptsize{$10$} && \scriptsize{$ 1.993 \cdot 10^{-3}$} && \scriptsize{$ 4.932 \cdot 10^{-3}$} && \scriptsize{$ 4.175 \cdot 10^{-4}$} && \scriptsize{$7.281 \cdot 10^{-3}$} \\
 
 \scriptsize{$10$} & \scriptsize{$100$} && \scriptsize{$7.232 \cdot 10^{-1}$} && \scriptsize{$ 5.114\cdot 10^{-3}$}  && \scriptsize{$7.976 \cdot 10^{-4}$} && \scriptsize{$7.183 \cdot 10^{-1} $}\\
 
 \scriptsize{$10 $} & \scriptsize{$234 $} && \scriptsize{$4.353$}   && \scriptsize{$ 1.176 \cdot 10^{-2} $}   &&  \scriptsize{$1.537 \cdot 10^{-3}$}  &&  \scriptsize{$4.366$}\\
 
  \scriptsize{$57$} & \scriptsize{$10$} && \scriptsize{$1.21\cdot 10^{-2}$} && \scriptsize{$7.594\cdot 10^{-3}$} && \scriptsize{$6.122\cdot 10^{-4}$} && \scriptsize{$2.039 \cdot 10^{-2}$}\\
 
 \scriptsize{$57$} & \scriptsize{$100$} && \scriptsize{$4.110$} && \scriptsize{$2.954 \cdot 10^{-2} $} &&  \scriptsize{$2.188 \cdot 10^{-2}$} && \scriptsize{$4.161$} \\
 
 \scriptsize{$57 $ } & \scriptsize{$234 $} && \scriptsize{$2.934 \cdot 10$} && \scriptsize{$2.983 \cdot 10^{-2} $}&& \scriptsize{$4.201 \cdot 10^{-3}$} && \scriptsize{$2.937 \cdot 10$}  \\ 
\bottomrule 
\end{tabular}\vspace{-1ex}
\caption{Computational cost of \textsc{ShapeIt}.}
\vspace{-4ex}
\label{table_comparison}
\end{wraptable}	
	
\paragraph{Wine dataset} This dataset~\cite{UCRArchive} consists of 111 traces, representing the food spectrograph  of two kinds of wine.  We consider only one class of wine data, containing 57 traces of length 234 samples (Fig.~\ref{fig:wine} shows one example).



By setting the maximum error threshold $\varepsilon_{\max}$ to $0.05$ (a little insight into how the learned specification varies depending on this value can be found in Table~\ref{table_threshold}), \textsc{ShapeIt} obtains an alphabet of four letters, each one describing a set of segments characterized by the values of slope, relative offset and duration reported in Fig.~\ref{fig:wine}.





The concatenation of letters $D$ and $A$ represents the peaks that appear in the shape (see Figure~\ref{fig:wine}), in which $D$ describes the rising part (with positive slope) and $A$ the decreasing one (with negative slope). Letter $C$ represents the approximately constant part of the trace that separates the two peaks, while $B$ describes the two extremes (they are both decreasing segments but less steep than the ones that come after the peaks' maxima).

In this particular application the values of slopes would be able to distinguish the different letters by their own: the intervals of slopes are indeed disjoint. The same happens for the relative offset but not for the duration.

\textsc{ShapeIt} generates an LSE specification (see Fig.~\ref{fig:wine}) that 
captures the two main peaks of the trace, but it is not able to recognize the little one that comes immediately after the first peak. The maximum error threshold $\varepsilon_{\max}$ should be reduced if one is interested in detecting also this little curve.


	    
	
\begin{wraptable}{r}{5cm}

\begin{tabular}{|r|c|c|c}
\toprule  
  \scriptsize{$\varepsilon_{\max}$}  &   \scriptsize{$\varphi$}   &  \scriptsize{\# clusters}  \\

\midrule 
\scriptsize{$0.05 $} & \scriptsize{$ B \cdot\  D \cdot  A \cdot C  \cdot D \cdot A \cdot B $} & \scriptsize{$4$} \\

\scriptsize{$0.1 $} & \scriptsize{$ F \cdot\  E \cdot  G \cdot I  \cdot H \cdot F $} & \scriptsize{$5$} \\

\scriptsize{$0.5 $} & \scriptsize{$ K \cdot (L +   M )$}  & \scriptsize{$3$}\\

\bottomrule  
\end{tabular}\vspace{-1ex}
\caption{Sensitivity w.r.t. $\varepsilon_{\max}$}
\label{table_threshold}
\vspace{-4ex}
\end{wraptable}
In Table~\ref{table_comparison}, we report the time (expressed in seconds) required by the tool to complete the three different phases:  segmentation ($t_s$), clustering ($t_c$) and automata learning ($t_l$). In the last column, $t_{\text{total}}$ summarizes the total time needed. Varying the number of traces and their lengths, we can observe that almost always the segmentation represents the most expensive part of the computation, while the clustering and the automata learning can be both considered negligible in terms of computation time. The only exceptions are the two cases in which the total number of traces is $1$ or $10$ with traces long only $10$: the values of $t_s$, $t_c$ and $t_l$ are comparable since the segmentation is very fast due to the low number of samples to approximate.

In Table \ref{table_threshold}, we compare the specifications learned varying the maximum error threshold $\varepsilon_{\max}$ from $0.05$ to $0.5$. The number of clusters does not decrease monotonically when increasing the maximum error allowed in the segmentation, while the specifications become shorter and therefore have less explanatory power.

 \vspace{-4ex}
\paragraph{Fish Data Set}
\begin{figure}[htp]
	    \centering
	    \includegraphics[width=\textwidth]{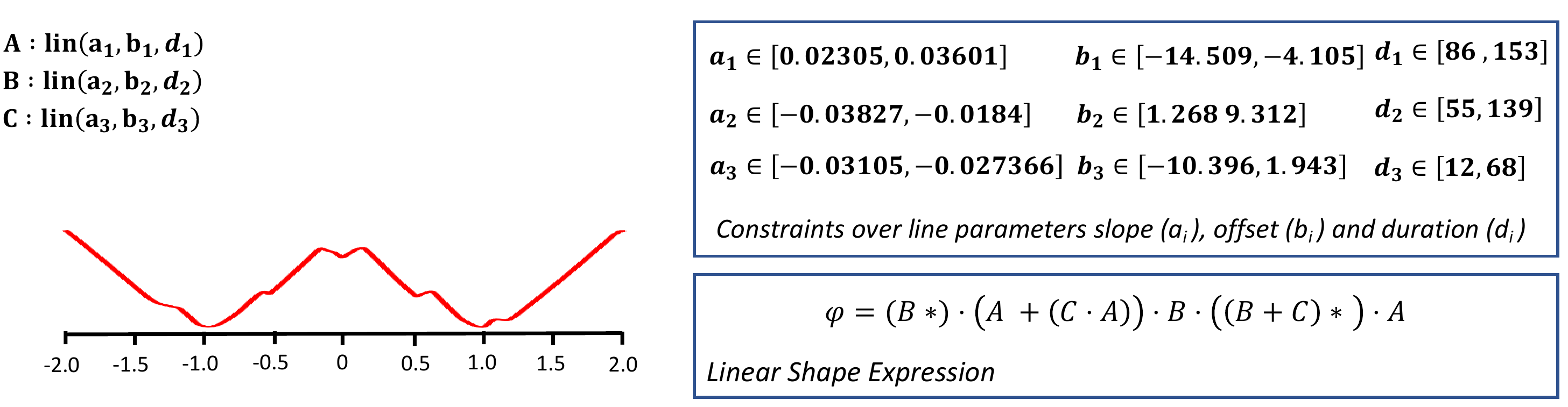}
	    \caption{(Left) Example of time series generated from the fish contour (see Fig. 20 of~\cite{Ken2006} for more details). (Right) Generated Linear Shape Expression.}
	    \label{fig3}
	    \vspace{-4ex}
	\end{figure}

This data set~\cite{UCRArchive} is composed by 350 time series representing the shape of seven different species of fishes (chinook salmon, winter coho, brown trout, Bonneville cutthroat, Colorado River cutthroat trout, Yellowstone cutthroat and mountain whitefish).  Starting from 50 images for each class, Lee \textit{ et al.} in~\cite{DahJye2008} generated the data set leveraging a novel technique that transforms the contour of the fish into a time series (see Fig.~\ref{fig3} on the left) using a turn-angle function as illustrated in~\cite{Ken2006}. Setting to 0.05 the maximum threshold error, with \textsc{ShapeIt} we are able to learn a specification (Fig.~\ref{fig3} on the right) from 26 shapes of the same species of fish, each one containing 463 samples. 









The concatenation of letters $B$ and $A$ represents the predominant shape in the traces: the triangular repeating behavior where, in particular, $A$ describes the rising part and $B$ the descending one (see Figure~\ref{fig3}). Letter $C$ is instead used to symbolize the noisy parts, both with positive and negative slope, that eventually separates these longer segments.
The choice operator ($+$) represents the possibility to have multiple symbols or expressions in different time series. 
Finally, the Kleene star ($*$) is used to indicate that a symbol or an expression can appear zero or more times. 

The learned specification provides insights about the relevant shapes in the time series data, displaying them in an human understandable language and therefore offering interpretability to the user. In this example, referring to the fish image in Fig.~\ref{fig3} on the left, we can associate the concatenation of letters $B$ and $A$ in the specification to the upper contour of the fish silhouette that is starting from the head and is ending with the tail. Since the same concatenation is then repeated in the specification, we can infer that the contour of the lower part of the fish is not significantly different from the upper one. Finally, letter $C$ can be interpreted as the presence of a big fin that interrupts the predominant lines described by letters $A$ and $B$.




	    
	

    \section{Conclusion and Future Work}
\label{sec:conc}

In this paper, we presented \textsc{ShapeIt}, a tool for mining specifications that describe the behaviors of CPS. The tool requires a set of real-valued signals generated by the system under study as input and it returns as output the specification that better summarize the properties of the traces in the form of linear shape expression. \textsc{ShapeIt} is structured in three phases: segmentation (approximating the traces with segments), abstraction and clustering (grouping lines with similar parameters) and automata learning (learning a DFA from words). Two additional values are needed as inputs to regulate the first two processes: a threshold expressing the maximum error allowed by the approximation and a threshold for the WCSS error to find an optimal number of clusters. We demonstrated the applicability of our tool over two different case studies (\textit{Wine} and \textit{Fish}) but other datasets are present in our repository. These data can be used as well to do experiments and gain confidence with \textsc{ShapeIt}.

As possible future works, we are interested in exploring and learning more general Shape Expressions (not necessarily linear ones), probably gaining explanatory power at the cost of an increasing computation time. We will also study how to automatize the tuning of the two thresholds required by the tool for the segmentation and the clustering phases. In this paper, the segmentation tool finds automatically the optimal number of segments to be used for the approximation, given a maximum error allowed. However it has already been developed to work in the other way round: receiving the number of required segments as input and then finding the approximation that provides the minimum error. It will be therefore interesting to exploit this feature to embed some domain knowledge (in the form of number of segments) in the specification mining process. A step forward will be adding the possibility to set constraints to the parameters of the lines. Finally, an other direction of work could be trying to generalize the tool in order to make it able to handle online processes instead of only offline ones.
	
	
	

	
	%
	%
	%
	
	\bibliographystyle{ieeetr}
	\bibliography{refs}

\begin{thebibliography}{10}

\bibitem{NenziSBB18}
L.~Nenzi, S.~Silvetti, E.~Bartocci, and L.~Bortolussi, ``A robust genetic
  algorithm for learning temporal specifications from data,'' in {\em Proc. of
  {QEST} 2018}, vol.~11024 of {\em LNCS}, pp.~323--338, Springer, 2018.

\bibitem{BBS14}
E.~Bartocci, L.~Bortolussi, and G.~Sanguinetti, ``Data-driven statistical
  learning of temporal logic properties,'' in {\em Proc. of {FORMATS}},
  pp.~23--37, 2014.

\bibitem{Wang2020}
F.~Wang, Z.~Cao, L.~Tan, and H.~Zong, ``Survey on learning-based formal
  methods: Taxonomy, applications and possible future directions,'' {\em IEEE
  Access}, vol.~8, pp.~108561--108578, 2020.

\bibitem{BartocciFMN18}
E.~Bartocci, T.~Ferr{\`{e}}re, N.~Manjunath, and D.~Nickovic, ``Localizing
  faults in {S}imulink/{S}tateflow models with {STL},'' in {\em {HSCC}},
  pp.~197--206, {ACM}, 2018.

\bibitem{JinDDS15}
X.~Jin, A.~Donz{\'{e}}, J.~V. Deshmukh, and S.~A. Seshia, ``Mining requirements
  from closed-loop control models,'' {\em {IEEE} TCAD}, vol.~34, no.~11,
  pp.~1704--1717, 2015.

\bibitem{BartocciMMMN19}
E.~Bartocci, N.~Manjunath, L.~Mariani, C.~Mateis, and D.~Nickovic, ``Automatic
  failure explanation in {CPS} models,'' in {\em {SEFM}}, vol.~11724 of {\em
  LNCS}, pp.~69--86, 2019.

\bibitem{BartocciDDFMNS18}
E.~Bartocci, J.~V. Deshmukh, A.~Donz{\'{e}}, G.~E. Fainekos, O.~Maler,
  D.~Nickovic, and S.~Sankaranarayanan, ``Specification-based monitoring of
  cyber-physical systems: {A} survey on theory, tools and applications,'' in
  {\em Lectures on Runtime Verification - Introductory and Advanced Topics},
  pp.~135--175, Springer, 2018.

\bibitem{NickovicQFMD19}
D.~Nickovic, X.~Qin, T.~Ferr{\`{e}}re, C.~Mateis, and J.~V. Deshmukh, ``Shape
  expressions for specifying and extracting signal features,'' in {\em Proc. of
  {RV}}, vol.~11757 of {\em LNCS}, pp.~292--309, 2019.

\bibitem{BartocciDGMNQ20}
E.~Bartocci, J.~Deshmukh, F.~Gigler, C.~Mateis, D.~Nickovic, and X.~Qin,
  ``Mining shape expressions from positive examples,'' {\em {IEEE} Trans.
  Comput. Aided Des. Integr. Circuits Syst.}, vol.~39, no.~11, pp.~3809--3820,
  2020.

\bibitem{UCRArchive}
Y.~Chen, E.~Keogh, B.~Hu, N.~Begum, A.~Bagnall, A.~Mueen, and G.~Batista, ``The
  {U}{C}{R} time series classification archive,'' July 2015.
\newblock \url{www.cs.ucr.edu/~eamonn/time\_series\_data/}.

\bibitem{Ken2006}
K.~Ueno, X.~Xi, E.~Keogh, and D.-j. Lee, ``Anytime classification using the
  nearest neighbor algorithm with applications to stream mining,'' in {\em
  Sixth International Conference on Data Mining (ICDM'06)}, pp.~623--632, 2006.

\bibitem{DahJye2008}
L.~Dah-Jye, J.~Archibald, R.~Schoenberger, A.~Dennis, and D.~Shiozawa, {\em
  Contour Matching for Fish Species Recognition and Migration Monitoring},
  vol.~122, pp.~183--207.
\newblock 2008.

\end{thebibliography}
	

\end{document}